\documentstyle[psfig,epsf,conf-X]{article}
\newcommand{\mincir}{\raise -2.truept\hbox{\rlap{\hbox{$\sim$}}\raise5.truept
\hbox{$<$}\ }}
\newcommand{\magcir}{\raise -2.truept\hbox{\rlap{\hbox{$\sim$}}\raise5.truept
\hbox{$>$}\ }}
\newcommand{\siml}{\raise -2.truept\hbox{\rlap{\hbox{$\sim$}}\raise5.truept
\hbox{$<$}\ }}
\newcommand{\simg}{\raise -2.truept\hbox{\rlap{\hbox{$\sim$}}\raise5.truept
\hbox{$>$}\ }}
\def\arcmin{\hbox{$^\prime$}}

\newcommand{\fl}{\,{\rm erg\,cm^{-2}s^{-1}}}
\newcommand{\lum}{\,{\rm erg\,s^{-1}}}
\begin{document} 
%
\null\vspace*{-3cm}\noindent \footnotesize{{\sf Proceedings of 
{\sl ``Large Scale Structure in the X-ray Universe"}, Santorini, Greece, Sept 1999}} 
\normalsize\vspace*{2cm}

\small
\heading{%
%
The Most Distant X-ray Clusters and the Evolution of their Space Density  
}
\par\medskip\noindent
\author{%
 P. Rosati$^{1}$, S. Borgani$^{2,3}$, R. Della Ceca$^4$,
A. Stanford$^5$, P. Eisenhardt$^6$, C. Lidman$^1$
}
\address{%
European Southern Observatory, D-85748 Garching bei M\"unchen, Germany
}
%
\address{%
INFN - Sezione di Perugia, c/o Dip. di Fisica, I-06121 Perugia, Italy
}
\address{%
INFN - Sezione di Trieste, c/o Dip. di Astronomia, I-34131 Trieste, Italy
}
\address{%
Osservatorio Astronomico di Brera, via Brera 28, I-20121 Milano, Italy
}
\address{%
Institute of Geophysics and Planetary Physics, Lawrence Livermore 
National Laboratory, P.O.Box 808, L-413, Livermore, CA 94550, USA
}
\address{%
Jet Propulsion Laboratory, California Institute of Technology, 
Mail Stop 169-327, Pasadena, CA 91109, USA
}

\begin{abstract}
We briefly review our current knowledge of the space density of
distant X-ray clusters as measured by several ROSAT serendipitous
surveys. We compare old and new determinations of the cluster
X-ray Luminosity Function (XLF) at increasing redshifts, addressing
the controversial issue of the evolution of its high end. We use 
complete subsamples, drawn from the ROSAT Deep Cluster Survey (RDCS), 
to quantify the statistical significance of the XLF evolution out to
$z\sim\! 1$. A consistent observational picture emerges
in which the bulk of the cluster population shows no
significant evolution out to $z\sim\! 1$, whereas the most luminous
systems ($L_X\magcir L^\ast_0\rm{[0.5-2\ keV]}\simeq 5\times 10^{44}\lum$)
were indeed rarer, at least at $z>0.5$,
in keeping with the original findings of the EMSS.  We also report on the
recent spectroscopic identification of four clusters in the RDCS lying 
beyond $z=1$, the most distant X-ray clusters known to date, which set an
interesting lower limit on the space density of clusters at $z >1$.
\end{abstract}
\section{Introduction}
Over the last five years, remarkable observational progress has been
made in constructing large samples of local and distant galaxy
clusters with the aim of quantifying the evolution of their space
density and providing the basis for follow-up studies of their
physical properties. The ROSAT satellite is largely responsible for
this progress, both with All-Sky Survey data and pointed observations,
which have been a gold mine for serendipitous discoveries.

About a thousand clusters have now been selected from the ROSAT All-Sky
Survey and several statistical complete subsamples have been used to
obtain a firm measurement of the local abundance of clusters
\cite{Ebe97,DeG99} and their spatial distribution (cf. B\"ohringer this
volume).  Serendipitous searches for distant clusters, selected as
extended X-ray sources in deep PSPC pointings
\cite{RDCS,Bur97,Jon98,Vik98}, have  boosted the number of known
clusters at $z>0.5$ by an order of magnitude, being just a few before
the ROSAT era. As we will show below, this recent work has complemented
the original Einstein Medium Sensitivity Survey (EMSS)
\cite{Gio90,EMSS}, and has corroborated its findings.

In this paper, we provide a brief update on our current knowledge of
the redshift dependent cluster X-ray Luminosity Function (XLF)
\cite{RosIAP} from results published over the last year.  The reader
is referred to the contributions of H.~Ebeling, I.~Gioia, L.~Jones,
A.~Vikhlinin in this volume for additional details and recent findings
on specific surveys.  We also report the most recent results from the
ROSAT Deep Cluster Survey (RDCS) which has allowed these studies to be
pushed beyond $z=1$ for the first time.

Unless otherwise stated, we assume $H_0=50$ km s$^{-1}$ Mpc$^{-1}$, $q_0=0.5$.
\begin{figure}[h]
\centerline{\vbox{
\psfig{figure=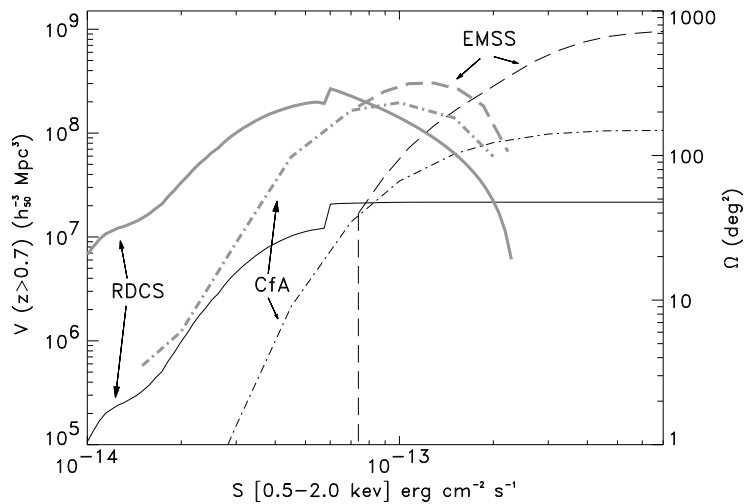,height=7cm}
}}
\caption{Sky coverage as a function of X-ray flux for 3 different
cluster surveys \cite{RDCS,Vik98,EMSS} (thin lines, right axis).  Thick
curves are the corresponding volumes (left axis) covered by each
survey at $z>0.7$ for a local $L_X^\ast$ cluster of $5\times 10^{44}\lum$.
}
\label{fig:vmax}
\end{figure}

\section{The Cluster X-ray Luminosity Function at $z<1$}

ROSAT distant cluster surveys \cite{RDCS,Bur97,Jon98,Vik98}, besides
employing different X-ray selection methods, have adopted different
strategies in terms of survey depth and solid angle.  In
Fig.\ref{fig:vmax}, we show the sky coverage of three surveys which
span a wide region of the solid angle--limiting flux plane, from the
large, shallow EMSS survey \cite{EMSS}, to the moderately deep 160
deg$^2$ CfA survey \cite{Vik98}, and the deep, small area (50 deg$^2$)
RDCS \cite{RDCS} (the other ROSAT surveys generally fill the space in
between).  This complementary coverage of the $\Omega-S$ plane has the
advantage of providing a better sampling of the XLF at different
redshifts when results from various surveys are combined.  As an
example, we also show in Fig.\ref{fig:vmax} the corresponding survey
volume which is covered at $z>0.7$ for an $L_X^\ast$ cluster.  This
illustrates the good sensitivity of the RDCS for detecting very
distant ``common" clusters, whereas a similar plot would show that the
EMSS explores a larger volume for the most luminous rare systems
($L>L^\ast$).

\begin{figure}
\centerline{\vbox{
\psfig{figure=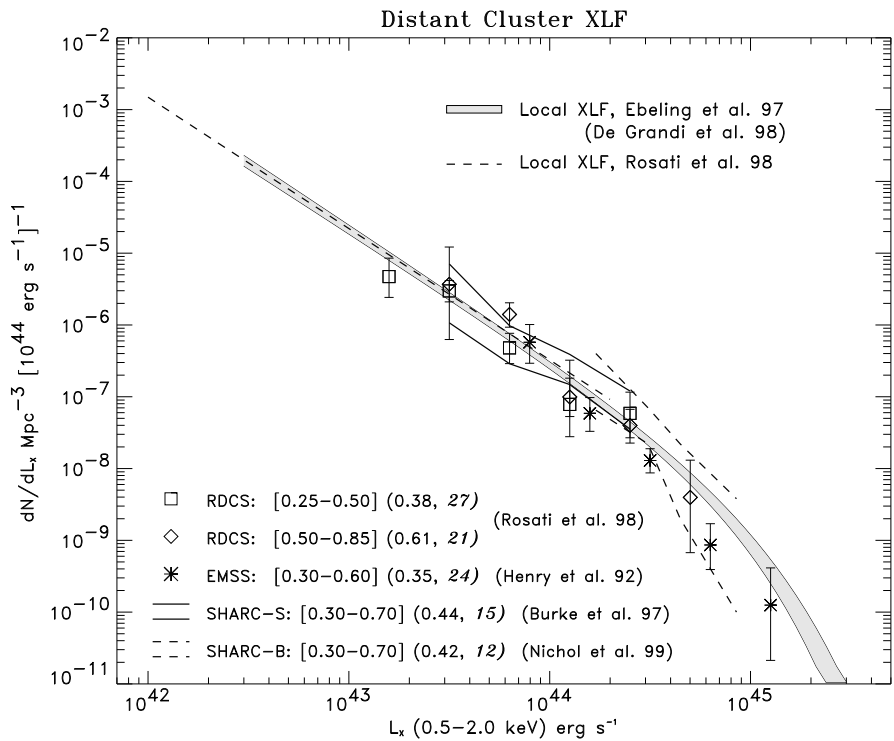,height=10cm,angle=90}
}}
\caption{Distant and local cluster XLFs from the literature. The number
in parenthesis are the median redshift and the number of clusters of
various samples in each redshift bin.}
\label{fig:xlf_all}
\end{figure}

In Fig.\ref{fig:xlf_all} we show several measurements of the cluster XLF 
that have been published to date. Sample sizes and median redshifts of 
each sample are 
also indicated. Based on these data, several groups have argued that
no significant evolution is observed in 
the space density of distant clusters
with $L_X[0.5-2\, \rm{keV}]\ \mincir 3\times 10^{44} \lum$
 \cite{Bur97,RDCS,Jon98,Vik98}. As demonstrated by the RDCS, this trend
persists out to $z\approx
0.8$. Measurements of the distant XLF at 
$L_X\ \magcir L^\ast_0 \simeq 5\times 10^{44} \lum$ are difficult with 
current samples, due to low number statistics. As a result, 
the evolution of the high end of the XLF has remained a hotly debated issue, 
ever since it was first reported in the EMSS \cite{Gio90,EMSS}.   
More recently, Vikhlinin et al. \cite{Vik98}
have confirmed the EMSS findings by comparing the observed number of 
very luminous systems with the no evolution prediction.
This result seems to be also in
agreement with a preliminary analysis of the {\sc Bright SHARC}
sample \cite{Nic99} (Fig.~\ref{fig:xlf_all}).

\subsection{Quantifying the XLF Evolution}

The binned representation of the XLF in Fig.\ref{fig:xlf_all} does not
provide a full picture of the space density evolution observed in a
given sample. For example, it fails to provide the statistical
significance of a possible departure from no evolution models
\cite{PC99}. The information contained in the RDCS can be more readily
recovered by analyzing the unbinned $(L_X,z)$ distribution with a
maximum-likelihood (ML) approach, which compares the observed cluster
distribution on the $(L_X,z)$ plane with that expected from a given
XLF model.

We characterize the cluster XLF as an evolving  Schechter function,\\
$\phi(L)=\phi_0 (1+z)^{A} L^{-\alpha}\exp(-L/L^\ast)$, with 
$L^\ast=L^\ast_0(1+z)^B$; where $A$ and $B$ are two evolutionary
parameters.  Different surveys find consistent values for the faint end 
slope $\alpha$, which is not observed to vary as a function of redshift
(Fig.~\ref{fig:xlf_all}). For the local XLF, we use here the measurement of
the BCS sample \cite{Ebe97}, i.e. $\alpha=1.85$, $L^\ast_0=5.7\times 10^{44}
\lum$, $\phi_0= 3.32\, (10^{-7} \rm{Mpc}^{-3}L_{44}^{\alpha-1})$.

\begin{figure}[t]
\centerline{\vbox{
\psfig{figure=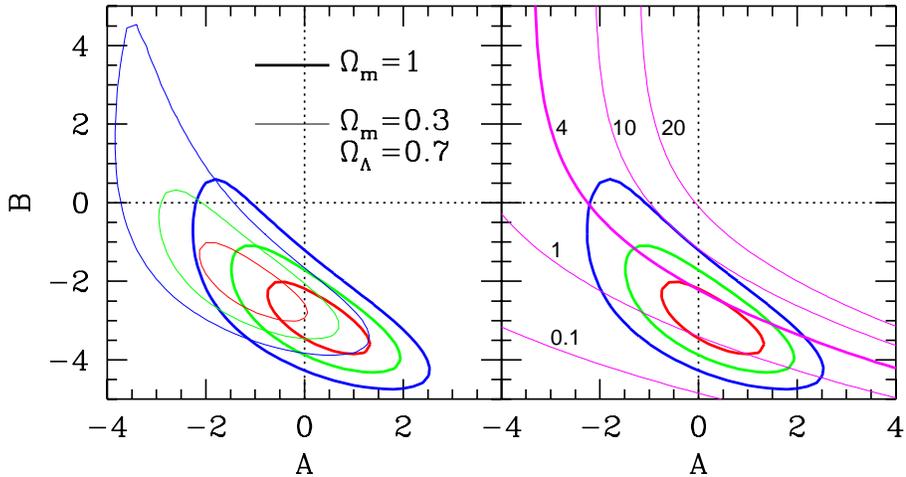,width=12.5cm}
}}
\caption{{\sl Left:} Confidence regions in the plane of the two 
evolutionary parameters A and B obtained by fitting the  function 
$\phi(L)=\phi_0 (1+z)^{A} L^{-\alpha}\exp(-L/L^\ast), \, 
L^\ast=L^\ast_0(1+z)^B$, to an RDCS subsample.
Maximum likelihood contours ($1\sigma$, $2\sigma$ and $3\sigma$ confidence 
levels) are plotted for two different  cosmologies. {\sl Right}: 
Loci of the A-B plane for which the corresponding XLF predicts 0.1, 1, 4,
10, 20 clusters at $z>1$ for the whole RDCS sample 
($F_{lim}=1\times 10^{-14}\fl$, sky coverage as in
Fig.\ref{fig:vmax}). Four clusters at $z>1$ have been confirmed in 
the RDCS to date.
}
\label{fig:like}
\end{figure} 

For this analysis, we use a complete flux limited sample
($F_{lim}=3.5\times 10^{-14}$ erg cm$^{-2}$ s$^{-1}$) 
of 81 spectroscopically confirmed RDCS clusters drawn
from 33 deg$^2$ ($z_{\rm{max}}=0.83$).  Observed flux errors are
included in the likelihood computation.  The resulting $1\sigma$,
$2\sigma$ and $3\sigma$ c.l. contours in the A-B plane are shown in
Fig.~\ref{fig:like}, for two different  cosmologies.  Best
fit values for the $\Omega_m=1$ case are $A=0.4^{+1.5}_{-1.8}$ and
$B=-3.0_{-1.2}^{+1.8}$ ($2\sigma$ errors). The no evolution model
($A=B=0$) is excluded at more than a $3\sigma$ confidence level, even
when the uncertainties of the local XLF are taken
into account.  The departure of our best fit model from the no-evolution 
scenario is due to the small number of observed clusters in the
RDCS at $z>\! 0.5$ with luminosities $L_X\ \magcir L^\ast_0$
compared to the no-evolution prediction. Interestingly, this effect
is barely significant with a slightly shallower sample
($F_{lim}=4\times 10^{-14}\fl$, 70 clusters).  
This evolutionary trend is similar to that observed in the EMSS
\cite{EMSS,Gio90}.

By excluding the most luminous clusters from our ML analysis, we find that  
there is {\sl no evidence of evolution} (with $2\sigma$ confidence level)
at luminosities $L_X\ \mincir 2\times 10^{44} \lum$, confirming 
previous results obtained with smaller samples. A redshift dependent 
inspection of the likelihood also shows that little can be said on the 
evolution of the high end of the XLF at $z\,\mincir 0.5$ with the current 
RDCS sample.

These findings lead to a consistent picture in which the comoving
space density of the bulk of the cluster population is  approximately
constant out to $z\sim\! 0.8$, but the most luminous ($L_X\ \magcir
L^\ast_0$), presumably most massive clusters were indeed rarer at high
redshifts.
Constraints on cosmological models based on this same RDCS sample are
discussed elsewhere (Borgani et al. this volume; \cite{BRTN99}).

\begin{figure}[t]
\centerline{
\psfig{figure=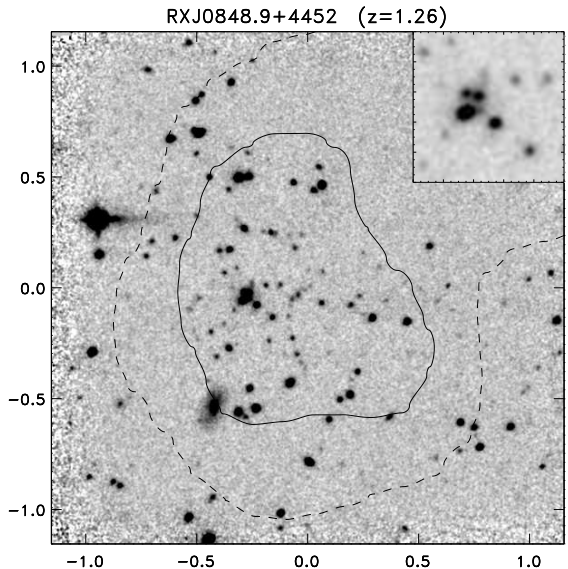,height=8cm}
}\vspace{-3truemm}
\caption{K-band image of RXJ0848.9+4452 with overlaid
 ROSAT-PSPC contours \cite{Ros99}. The inset is a blowup of the core. 
The scale is in arcmin.
}
\label{fig:c0848}
\end{figure}

\section{Beyond $z=1$}
An inspection of Fig.~\ref{fig:vmax} indicates that the RDCS probes
an appreciable volume at high redshifts. The maximum sensitivity for
clusters at $z\,\magcir 1$ is reached at fluxes below $3\times
10^{-14}$ and for luminosities $L_X\approx 2\times 10^{44} \lum$. The
discovery of the first X-ray selected cluster (RXJ0848.9+4452, 
Fig.~\ref{fig:c0848}) at 
$z=1.26$ in the RDCS \cite{Ros99} has confirmed these expectations.
Deep near-IR imagery and optical spectroscopy with
Keck/LRIS were required to secure this identification.
This system has $L_X\simeq 1.5\times
10^{44}\lum$ (in rest frame [0.5-2 keV] band) and is found to lie
only $4.2\arcmin$ away ($5.0\, {\rm h}^{-1}_{50}$ comoving Mpc) from
an IR selected cluster previously discovered by
Stanford et al. \cite{S97} at $\langle z\rangle =1.273$, also known to
be X-ray luminous with half the $L_X$ of RXJ0848.9.  This is, most
likely, the first example of a high-redshift supercluster consisting
of two separate systems in an advanced stage of collapse.  Scheduled
Chandra and XMM observations of this field should provide important
information on the temperature and metal enrichment of 
their intra-cluster media.

\begin{figure}[t]
\centerline{
\psfig{figure=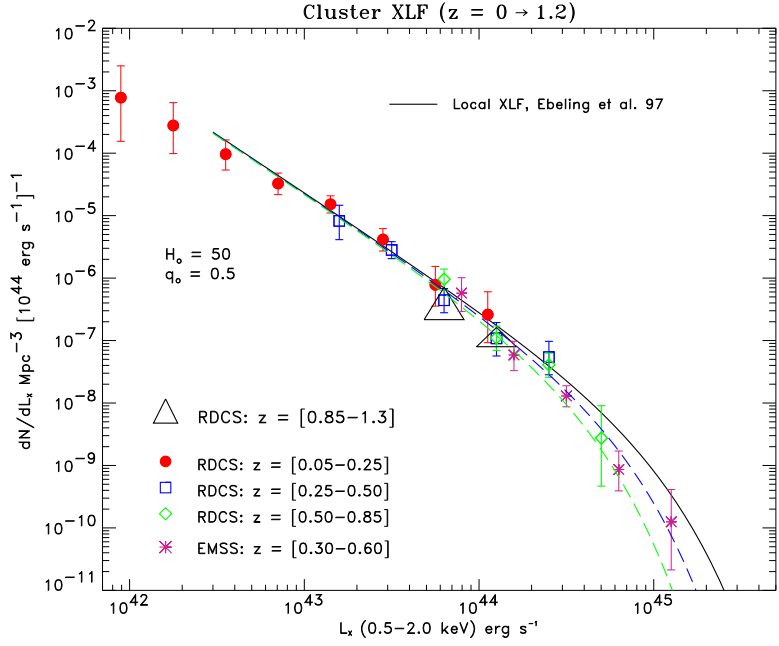,height=8.7cm,angle=90}
}\vspace{-2truemm}
\caption{The best determination to date of the cluster XLF out to
$z\simeq 1.2$. Data points at $z<0.85$ are derived from a complete
RDCS sample of 103 clusters over 47 deg$^2$, with $F_{lim}=3\times 10^{-14}
\fl$. The triangles represent  {\it a lower limit} (due to incomplete
optical identification) to the cluster space density obtained from a
sample of 4 clusters with $\langle z\rangle=1.1$ and
with $F_{lim}=1.5\times 10^{-14} \fl$.
Long dash curves are Schechter best fit XLFs for $z=0.4$ and
$z=0.6$, according to the model of fig.~\ref{fig:like}.
}
\label{fig:xlf_new}
\end{figure}

Recently, two additional faint RDCS candidates have been spectroscopically 
confirmed, using Keck/LRIS and VLT/FORS, as clusters at $z=1.10$ and
$z=1.23$. It should be stressed that these clusters, with $F_X\approx
2\times 10^{-14}\fl$, are low surface brightness ``fluctuations'' in
PSPC images, and therefore the fraction of spurious candidates can be
significant at these low fluxes. Moreover, the spectroscopic follow-up
work of such distant clusters is particularly time consuming, even 
with 8-10 meter class telescopes.
Although the optical identification is not yet complete at these faint
flux levels, the high-$z$ tail of the RDCS can be used to set an
interesting lower limit to the space density of clusters at $\langle
z\rangle=1.1$.  This is shown in Fig.~\ref{fig:xlf_new}, where an
extended RDCS sample has been used to obtain the best estimate of the
XLF of distant clusters out to $z\approx 1.2$. Such a sample contains
107 clusters drawn from a 47 deg$^2$ area, with 8 clusters at $z>0.8$.
We also plot in Fig.~\ref{fig:xlf_new} the best fit XLF model
described above, at $z=0.4$ and $z=0.6$. 

To better understand the constraints that these newly-identified
high redshift clusters set on the XLF evolution, we have plotted in
Fig.~\ref{fig:like} (right) the loci of the A-B plane for which the
corresponding XLF, $\phi_{[A,B]}(L,z)$, predicts
$0.1,\,1,\,4,\,10,\,20$ clusters at $z>1$, for the entire RDCS sample.
About 20 clusters would have to be identified in the no-evolution
scenario.  This seems very unlikely, unless the sample is severely
incomplete at faint fluxes. Given that 4 clusters have already
been discovered, the portion of the A-B plane which is allowed by this
preliminary analysis suggests that the evolution is still rather mild at 
$z\,\magcir 1$, at luminosities just above $10^{44}\lum$.

The next obvious step in the effort to understand cluster
formation and evolution is to push the cluster (or proto-cluster)
search out to even higher
redshifts, namely out to $z\sim\!3$ where the signature 
of large scale stucture has
already been unveiled \cite{Ste99}. Finding clusters around high-$z$
AGN is a viable method (e.g.\cite{Fab96,Dic99,Car98}), although not suitable 
for assessing the cluster abundance.
Serendipitous searches with Chandra and XMM will of course be actively
pursued, but it will take several years to build large enough
survey areas, and furthermore, the spectroscopic follow-up of cluster
candidates at $z>1.3$ may turn out to be too difficult with existing 
telescopes.  
While the short-term prospects for exploring the era at $1.5\,\mincir z\,
\mincir 2.5$ may appear somewhat bleak, it should be kept in mind that
earlier this decade many theorists and observers were convinced that clusters
at $z>1$ were either out of reach, or did not exist.

\vfill
\end{document}